\documentclass[%
 amsmath,amssymb,amsfonts,
 aps,
 pre,
 twocolumn,
 longbibliography
]{revtex4-2}

\usepackage [english]{babel}

\usepackage{graphicx}
\usepackage[hidelinks]{hyperref}
\usepackage[export]{adjustbox}

\begin{document}

\title{Performance evaluation of variational quantum eigensolver and\\ quantum dynamics algorithms\\
on the advection-diffusion equation}

\author{A. Bar\i\c{s} \"Ozg\"uler}
\email{baris\_ozguler@berkeley.edu}
\affiliation{%
Haas School of Business, University of California, Berkeley, California 94720, USA
}%

\makeatletter
\renewcommand*{\Dated@name}{}
\makeatother
\date{8 June 2026}

\begin{abstract}
Near-term quantum algorithms are a promising route to solving partial differential equations, but gauging their true potential requires separating algorithmic performance from sampling and hardware noise. We benchmark a ground-state variational quantum eigensolver (VQE), cast as a variational quantum linear solver, against the Trotterization, variational quantum imaginary time evolution, and adaptive variational quantum dynamics simulation methods applied to the one-dimensional advection-diffusion equation in the recent quantum-dynamics study by Alipanah \emph{et al.}~[Phys.\ Rev.\ Res.\ \textbf{7}, 043318 (2025)] at matched grid and problem size. On a noiseless state-vector simulator the $N=4$ VQE drives the final-time infidelity to a numerical floor ($\sim\!10^{-14}$) once the depth reaches $L\approx5$, an \emph{algorithmic ceiling} set by exact expectation values. Evaluating the same solver with a finite number $S$ of measurement shots, still without hardware noise, makes the infidelity sampling limited, following $1-f\approx c/S$ (a best-case readout-sampling estimate, with the solution's signs assumed known), providing a regime-matched comparison with the shot-based emulator of Alipanah \emph{et al.}\ and explaining the gap to their noisy hardware runs ($>10^{-1}$). The benchmark thus decomposes the near-term error budget into algorithmic, sampling, and hardware contributions, with a matched-depth resource comparison. The formulation applies without modification across $N=4,5,6$ qubits and to a two-dimensional (eight-qubit, $16\times16$) problem evolved to $t=1$, where the state-vector VQE holds a $\sim\!10^{-7}$ algorithmic-ceiling infidelity against the sampling-limited $\sim\!10^{-5}$ of the corresponding shot-based simulation, a difference of measurement regime rather than algorithmic superiority.
\end{abstract}

\maketitle

\section{Introduction}
\label{sec:sec1}

Quantum computing has emerged as a leading paradigm to speed up the solution of high-dimensional partial differential equations (PDEs), which arise in fluid dynamics, finance, materials science, and machine learning \cite{GH19_National,givi2020quantum,alexeev2021prxq,daley2022nat, Succi_2023,metcalfe2023aiaa,felix2023weather,tennie2025quantum,bmw2024applications,Succi24, sanavio2024explicit, yeung2024rpp, sanavio2025carleman, zecchi2025improved, xu2025improved}. Classical methods often suffer from exponential scaling in memory or runtime, especially for fine-resolution simulations. Recent advances in quantum algorithms, particularly variational quantum algorithms (VQAs) and quantum dynamics solvers, offer new tools for addressing these challenges on noisy intermediate-scale quantum (NISQ) hardware. Recent studies have begun to formalize quantum benchmarking approaches and metrics in the presence of noise, fostering a more systematic evaluation of quantum algorithms in near-term hardware environments \cite{proctor2022measuring, bharti2022rmp, lubinski2023application, lall2025review}.

One especially challenging domain is the numerical solution of PDEs describing transport phenomena (e.g., in fluid dynamics, heat transfer, and finance). Classical approaches often face high memory and runtime costs, particularly in higher dimensions or when fine mesh resolutions are required \cite{LNBG20,Wang2015TowardsHA}. Quantum algorithms, in principle, may alleviate these overheads by encoding the solution into amplitudes of a quantum state whose dimension grows exponentially with the number of qubits \cite{reiher2017elucidating,von2021quantum,jaksch2023variational,santagati2024drug,meng2023prr,brearly2024pra, sanavio2024explicit, huang2024pite}. 

Before fault-tolerant quantum hardware becomes available, one may still explore quantum algorithms for solving PDEs on today's NISQ devices \cite{preskill2018quantum,bharti2022rmp,callison2022pra,osaba2024production, wright2024noisy}. VQAs have emerged as a leading candidate for NISQ-era simulations due to their potential resilience against noise \cite{cerezo2021nat,BravoPrieto2023vqls,blekos20241pr}. These hybrid quantum-classical schemes adopt parameterized circuits and iteratively update the parameters to minimize a suitable cost function, enabling approximate solutions to quantum chemistry, condensed matter, and PDE-based problems on small noisy hardware \cite{jaksch2023variational,rigas2022vqa,guseynov2023pra,ali2023pra,guzman2024pra,sarma2024pra,li2023variational,pool2024prr,liu2024ocean,hunout2024variational,song2025incompressible,choi2024vqa,margarit2024engineering,bosco2024variational,syamlal2024variational,turati2024variational,surana2024variational,arora2024finelem,ingelmann2024advdif,bengoechea2024vqa,kocher2024nonlinear_schrod,over2025boundary,gnanasekaran2024variational,fathi2024hybrid,amaro2024fokker, li2023benchmarking, ozguler2022numerical, alam2022quantum, xu2022neural, leong2022variational}. 
Compared to direct Hamiltonian simulation methods like Trotterization for qubits and qudits \cite{motta2020determining, ogunkoya2024qutrit, ozguler2024qudit, kamakari2022digital, sun2021prxq}, VQAs typically require fewer gates, although they involve classical optimization loops prone to problems such as barren plateaus \cite{mcclean2018barren,wang2021noise,larocca2024review,cunningham2025investigating}.

Recently, Alipanah \emph{et al.}\ solved
the one-dimensional advection-diffusion equation on IBM quantum hardware 
via three quantum-dynamics algorithms: Trotterization, variational quantum imaginary time evolution (VarQTE), and adaptive variational quantum dynamics simulation (AVQDS) 
\cite{alipanah2025quantum}. On a noiseless shot-based emulator, they 
achieved final infidelities on the order of $10^{-5}$, demonstrating 
the algorithms' intrinsic capabilities when measurement outcomes
are sampled but no hardware noise is present. However, in actual 
runs on a physical quantum processor, those infidelities increased
to $>10^{-1}$, underscoring the impact of real
hardware noise. In addition to their hardware-based one-dimensional (1D) studies, Alipanah \emph{et al.}~\cite{alipanah2025quantum}\ also explored a 2D advection-diffusion system at the simulator level. Although their noise-free simulations yielded encouraging results, the required circuit sizes remain beyond the capabilities of current NISQ processors, and the authors conclude that a hardware demonstration is currently infeasible. This further underscores the importance of establishing noise-free benchmarks, such as those provided by idealized state-vector simulations, to gauge the intrinsic performance of quantum PDE solvers before pushing toward higher-dimensional cases on real devices.

This study is a systematic \emph{performance evaluation} of near-term quantum solvers for the
advection-diffusion equation: We benchmark a variational quantum eigensolver (VQE) approach for the PDE studied by Alipanah \emph{et al.}~\cite{alipanah2025quantum},
beginning in an idealized (noise-free) state-vector regime and then adding finite sampling. By comparing to
their results, we disentangle purely \emph{algorithmic} performance (how well an ansatz-based solver
represents the PDE solution) from sampling and hardware-induced errors. Our contributions are as follows: (i) a clear
single-register formulation of PDE time stepping as a variational quantum linear solver, in which each
forward-Euler step is the zero-eigenvalue ground state of an explicit positive-semidefinite Hamiltonian,
with a short proof and an explicit cost function; (ii) a head-to-head performance evaluation of this
ground-state solver against the quantum-dynamics methods (Trotter, VarQTE, and AVQDS) studied by Alipanah
\emph{et al.}, on the \emph{same} PDE, grid, and problem size,
complementing prior quantum-PDE benchmarking efforts \cite{song2023qpde}; (iii) a noise-free state-vector
ceiling (final-time infidelity $\sim\!10^{-14}$, a numerical floor reached by depth $L\approx5$ on four qubits) together with a noiseless
\emph{shot-based} study showing the infidelity is sampling-limited to $\approx c/S$, bridging the
state-vector ceiling and the $>10^{-1}$ hardware errors of Ref.~\cite{alipanah2025quantum}; and
(iv) a matched-depth, per-time-step resource comparison. The outcome is a decomposition of the near-term
error budget into algorithmic, sampling, and hardware contributions, without asserting that any single
method is uniformly superior.

The VQE optimization used here employs exact (state-vector) expectation values and analytic gradients in place of shot-estimated partial derivatives, so it is not yet directly hardware-deployable. Section~\ref{sec:shotnoise} quantifies the sampling-limited regime directly; a fully shot-based optimization (estimating the cost and its gradients from measurements, and subject to an additional $1/\sqrt{N_s}$ cost-estimation floor) is the natural remaining step toward hardware. This baseline helps identify the true algorithmic potential, clarifies the sources of error, and points to the next steps in bridging the gap to real NISQ performance. A fully shot-based loop would also have to contend with trainability: The $1/\sqrt{N_s}$ cost-estimation floor compounds with any flattening of the variational landscape (barren plateaus~\cite{mcclean2018barren,larocca2024review}) as the register and ansatz grow. For the small $N$ and shallow, warm-started circuits used here these effects are mild, but we expect them to become a binding constraint, alongside the expressivity wall (Sec.~\ref{sec:Nsweep}), at larger $N$.

The rest of the paper is organized as follows. 
Section~\ref{sec:sec2} reviews the one-dimensional advection-diffusion PDE and discretizes it via finite differences. 
Section~\ref{sec:sec3} summarizes classical DNS and outlines how the PDE can be solved by either (i) the VQE-based linear-systems approach or (ii) quantum dynamics schemes (Trotter, VarQTE, and AVQDS). 
Section~\ref{sec:sec3-1} then presents our VQE results, including comparisons of gate counts to Ref.~\cite{alipanah2025quantum}, infidelities versus time, and the effect of circuit depth. 
Finally, Sec.~\ref{sec:Conclusions} concludes with an outlook on scaling to larger PDEs, the importance of noise analysis, and applications in finance and fluid engineering.

\section{Advection-Diffusion PDE and Quantum Encoding}
\label{sec:sec2}

We consider the dimensionless 1D advection-diffusion equation:
\begin{equation}
    \frac{\partial C}{\partial t} 
    + \frac{\partial C}{\partial x} 
    \;=\;
    \frac{1}{Pe}\,\frac{\partial^2 C}{\partial x^2},
    \label{eq:eq1_ade}
\end{equation}
with P\'eclet number $Pe = \tfrac{LU}{\Gamma}$ quantifying advective versus diffusive transport. Here $C(x,t)$ is a scalar field, $0 \le x \le 1$ is the domain (normalized by length $L$), and $t\ge 0$. Periodic boundary conditions are assumed. A second-order central finite-difference scheme is applied in space:
\begin{align}
\label{eq:eq2_ade_fd}
\frac{\partial C_i}{\partial t} + \frac{C_{i+1} - C_{i-1}}{2\,\Delta x}
&=\frac{1}{Pe}\,\frac{C_{i+1} - 2\,C_i + C_{i-1}}{\Delta x^2},
\end{align}
with $i=0,\dots,2^N-1$ and $\Delta x = \tfrac{1}{2^N}$. We collect $C(x_i)$ into a $2^N$-component vector $|C\rangle$ for quantum encoding, so that the $i$-th amplitude corresponds to $C(x_i)$. For $N=4$, we have $16$ spatial points.

Rewriting \eqref{eq:eq2_ade_fd} in vector form,
\begin{equation}
    \frac{\partial |C\rangle}{\partial t} \;=\;\hat{A}\,|C\rangle,
    \label{eq:eq3_ade}
\end{equation}
\begin{equation}
    \hat{A} \equiv \frac{1}{Pe \, \Delta x}\begin{pmatrix}
    b & c & 0 & 0 & 0 & \dots & 0 & d \\
    d & b & c & 0 & 0 & \dots & 0 & 0 \\
    0 & d & b & c & 0 & \dots & 0 & 0 \\
    0 & 0 & d & b & c & \dots & 0 & 0 \\
    \vdots & \vdots & \vdots & \vdots & \vdots & & \vdots & \vdots \\
    c & 0 & 0 & 0 & 0 & \dots & d & b
\end{pmatrix}, 
\label{eq:eq4}
\end{equation}
where $\quad b = -\frac{2}{\Delta x}, \quad c = \frac{1}{\Delta x} - \frac{Pe}{2}, \quad d = \frac{1}{\Delta x} + \frac{Pe}{2}$, and $\hat{A}$ is a $2^N\times 2^N$ operator (non-Hermitian, due to advection). We assume a normalized state vector $|C\rangle$ with periodic boundary conditions.

\section{Methodology: Classical (DNS), VQE, and Quantum Dynamics Approaches}
\label{sec:sec3}

\begin{figure*}[!htbp]
\centering
\includegraphics[width=\linewidth]{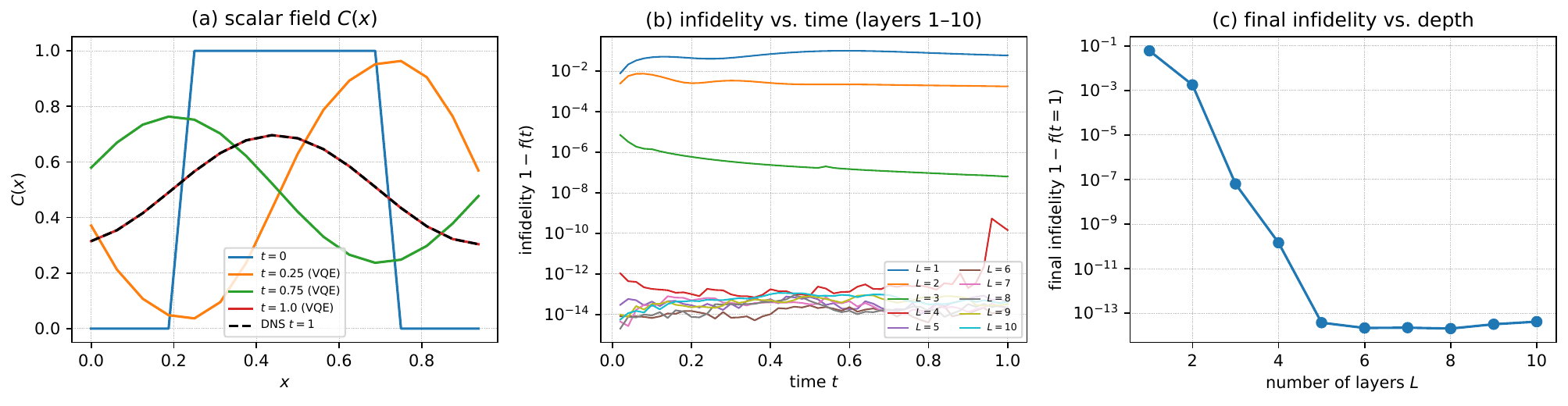}
\caption{\textbf{(a)} Scalar field $C(x)$ reconstructed from VQE outputs at various time points ($t = 0$, $0.25$, $0.75$, $1.0$), alongside the corresponding DNS result at $t = 1.0$ (dashed black line). \textbf{(b)} Infidelity $1 - f$ between the VQE and DNS solutions vs time for circuit depths $L=1$--$10$. With a converged optimizer the infidelity reaches a numerical floor ($\sim\!10^{-14}$) by $L\approx5$, since the $4$-qubit real-amplitudes ansatz already has enough parameters ($\geq15$) to represent any real $16$-dimensional state. \textbf{(c)} Final-time infidelity ($t=1$) vs number of layers.}
\label{fig:infidelities}
\end{figure*}

\subsection{Classical direct numerical simulation}
A simple forward-Euler scheme updates $|C_{k}\rangle \mapsto |C_{k+1}\rangle$ in time:
\begin{equation}
  |C_{k+1}\rangle 
  \;=\;
  \bigl(I + \Delta t\,\hat{A}\bigr)\,|C_k\rangle,
  \label{eq:euler_dns}
\end{equation}
for $k=0,1,\dots,N_{\mathrm{steps}}-1$, with time step $\Delta t=t_{\max}/N_{\mathrm{steps}}$. This direct numerical simulation (DNS) provides a stable reference solution, provided $\Delta t$ and $Pe$ satisfy stability constraints. In the results below, we label this classical vector as $|C_{\mathrm{DNS}}(t)\rangle$ and use it as ground truth to measure infidelities. The eigenvalues of $\hat A$ have negative real parts set by diffusion ($\sim-k^2/Pe$) and imaginary parts set by advection; explicit stability of \eqref{eq:euler_dns} requires $|1+\Delta t\lambda|\le1$, which for the diffusive part reduces to the familiar bound $\mathrm{CFL}_{\mathrm{diff}}=\Delta t\,4^{N}/Pe\lesssim\tfrac12$ used in Sec.~\ref{sec:Nsweep}. We adopt explicit Euler as the reference scheme because it makes each step an exact linear-system solve (Sec.~\ref{sec:vqe_form}); implicit or higher-order integrators such as Crank-Nicolson would relax this step-size restriction and fit the same VQLS template, with $I+\Delta t\hat A$ replaced by the appropriate step operator, at the cost of a less sparse operator to decompose.

\subsection{VQE-based linear-system formulation}
\label{sec:vqe_form}
To implement the time step \eqref{eq:euler_dns} on a quantum device, we recast it as a
linear-system problem and solve that problem variationally, following the variational
quantum linear solver (VQLS) framework \cite{BravoPrieto2023vqls}. Treating the known current state
$|C_k\rangle$ as the right-hand side and the next state $|C_{k+1}\rangle$ as the unknown, the
forward-Euler update \eqref{eq:euler_dns} is equivalent to the linear system
\begin{equation}
  \mathcal{A}\,|x\rangle = |b\rangle,
  \qquad
  \mathcal{A} \equiv \bigl(I+\Delta t\,\hat{A}\bigr)^{-1},
  \label{eq:linsys}
\end{equation}
with unknown $|x\rangle=|C_{k+1}\rangle$ and right-hand side $|b\rangle=|C_k\rangle$. Here
$\mathcal{A}=(I+\Delta t\hat{A})^{-1}$ is a $2^N\times2^N$ operator ($I+\Delta t\hat{A}$ is invertible
whenever $-1/\Delta t$ is not an eigenvalue of $\hat{A}$, which holds for the stable steps used here),
$|b\rangle=|C_k\rangle$ is the normalized current solution, and
$|x\rangle=|C_{k+1}\rangle$ is the normalized next solution. Both $|x\rangle$ and $|b\rangle$ are
genuine $N$-qubit quantum states, so the encoding uses $N$ qubits and a Hilbert space of dimension
$2^N$ (for $N=4$, $16$ amplitudes), the \emph{same} size as the spatial grid; no register doubling
is required.

Following the VQLS construction, the solution of \eqref{eq:linsys} is obtained as the ground state of a
positive semi-definite Hamiltonian, as the following proposition makes precise.

\emph{Proposition.} Let $\mathcal{A}=(I+\Delta t\hat{A})^{-1}$ be invertible and $|b\rangle=|C_k\rangle$
be normalized ($\langle b|b\rangle=1$). Then
\begin{equation}
  H_k \;=\; \mathcal{A}^\dagger\bigl(I-|b\rangle\langle b|\bigr)\mathcal{A},
  \qquad b=C_k ,
  \label{eq:Hk}
\end{equation}
is positive semi-definite, and its zero-eigenvalue ground state is unique and equals the normalized next
time step $|C_{k+1}\rangle$.

\emph{Proof.} For any normalized $|\psi\rangle$,
\begin{equation}
  \langle\psi|H_k|\psi\rangle
  = \big\|\mathcal{A}|\psi\rangle\big\|^2 - \big|\langle b|\mathcal{A}|\psi\rangle\big|^2 \;\ge\; 0 ,
  \label{eq:cost}
\end{equation}
by the Cauchy-Schwarz inequality (using $\langle b|b\rangle=1$), so $H_k\succeq0$. Equality (i.e.\ a
zero of the cost, hence a vector in $\ker H_k$) holds if and only if $\mathcal{A}|\psi\rangle\parallel
|b\rangle$, i.e.\ $|\psi\rangle\propto\mathcal{A}^{-1}|b\rangle=(I+\Delta t\hat{A})|C_k\rangle\propto
|C_{k+1}\rangle$. Since $\mathcal{A}$ is invertible, $\ker H_k=\mathrm{span}\{\mathcal{A}^{-1}|b\rangle\}$
is one dimensional, so the zero-eigenvalue ground state is unique (up to global phase) and equals the
normalized $|C_{k+1}\rangle$. $\quad\square$

We verified numerically that the smallest eigenvalue of $H_k$ vanishes to machine precision, with an
order-unity spectral gap ($\approx1$) to the next eigenvalue, and that its eigenvector reproduces
\eqref{eq:euler_dns}.

We obtain this ground state with a VQE
\cite{jaksch2023variational,ali2023pra,BravoPrieto2023vqls} by minimizing the cost
\eqref{eq:cost}, $C(\theta)=\langle\psi(\theta)|H_k|\psi(\theta)\rangle$, over the parameters
$\theta$ of a hardware-efficient ansatz $|\psi(\theta)\rangle$ (Sec.~\ref{sec:sec3-1}); an
equivalent normalized cost is $\hat{C}(\theta)=1-|\langle b|\mathcal{A}|\psi(\theta)\rangle|^2/
\langle\psi(\theta)|\mathcal{A}^\dagger\mathcal{A}|\psi(\theta)\rangle\in[0,1]$. Because both the
cost and the fidelity are invariant under $|x\rangle\to e^{i\varphi}|x\rangle$ and under rescaling,
the system is solved projectively, $\mathcal{A}|x\rangle\propto|b\rangle$: The proportionality
constant equals the per-step norm change $\|(I+\Delta t\hat{A})|C_k\rangle\|$, which we track
classically by renormalizing at each step. Repeating for $k=0,\dots,N_{\mathrm{steps}}-1$ yields
the full time evolution. An equivalent block (register-doubling) formulation and a fully worked
$N=4$ example are given in Appendix~\ref{app:formulation}.

\subsection{Quantum dynamics approaches: Trotter, VarQTE, and AVQDS}
An alternative is to \emph{directly} simulate 
\begin{equation}
\label{eq:ade_imag_time}
  \frac{d}{d\beta}\,|C\rangle(\beta) 
  = -\,i\,\hat{A}\,|C\rangle(\beta),
  \quad
  \beta = i\,t,
\end{equation}
which recasts the PDE in a Schr\"odinger-like equation with (complex) $\beta$. Trotterization decomposes $e^{-i\,\beta\hat{A}}$ into exponentials of Pauli strings, typically requiring many gates \cite{kumar2024generalising,guseynov2024explicit}. Variational time-evolution methods (VarQTE and AVQDS) replace the full operator with a parameterized ansatz and update angles in time \cite{mcardle2019variational,yuan2019theory,yao2021adaptive}, significantly lowering the gate count. Reference~\cite{alipanah2025quantum} reported that Trotter circuits for $N=4$ can require $\sim 10^5$ gates in total ($\sim2\times10^{4}$ of them entangling), whereas VarQTE and AVQDS require $\sim 10^2$ gates but still suffer from hardware noise. 

\section{VQE Simulation Results}
\label{sec:sec3-1}

We solve the advection-diffusion equation \eqref{eq:eq1_ade} with P\'eclet number $Pe=32$, on the
domain $x\in[0,1)$ with periodic boundaries, and $N=4$ qubits, so that there are $2^N=16$ grid points
with spacing $\Delta x=1/2^N=1/16=0.0625$ and $x_i=i\,\Delta x$, $i=0,\dots,15$. We use a \emph{box}
initial condition, $C(x_i,0)=1$ for $4\le i\le 11$ (i.e.\ $0.25\le x<0.75$) and $0$ otherwise, which
after normalization is
\begin{equation}
  |C_0\rangle=\tfrac{1}{\sqrt8}\,(0,0,0,0,1,1,1,1,1,1,1,1,0,0,0,0)^{\!\top},
  \label{eq:ic}
\end{equation}
with amplitude $1/\sqrt8\approx0.354$ on the eight central grid points. We use a time step
$\Delta t=0.002$ and integrate until $t=1.0$ ($N_{\mathrm{steps}}=500$). At each step we build $H_k$
from \eqref{eq:Hk} and minimize the cost \eqref{eq:cost} with the gradient-based \texttt{L-BFGS-B}
optimizer (up to $5000$ iterations per step, with parameters warm-started from the previous step);
gradient-free optimizers (COBYLA, SPSA) reach the same solutions at higher iteration cost. The cost is
evaluated from the full state-vector (no shots); shot-based evaluation is studied in
Sec.~\ref{sec:shotnoise}. The circuit ansatz is a real-amplitudes ($R_y$+CNOT) design with variable
depth (\emph{layers}). Figure~\ref{fig:infidelities} summarizes the resulting solutions and infidelities.

\begin{figure*}[!htbp]
\centering
\includegraphics[scale=0.53,left]{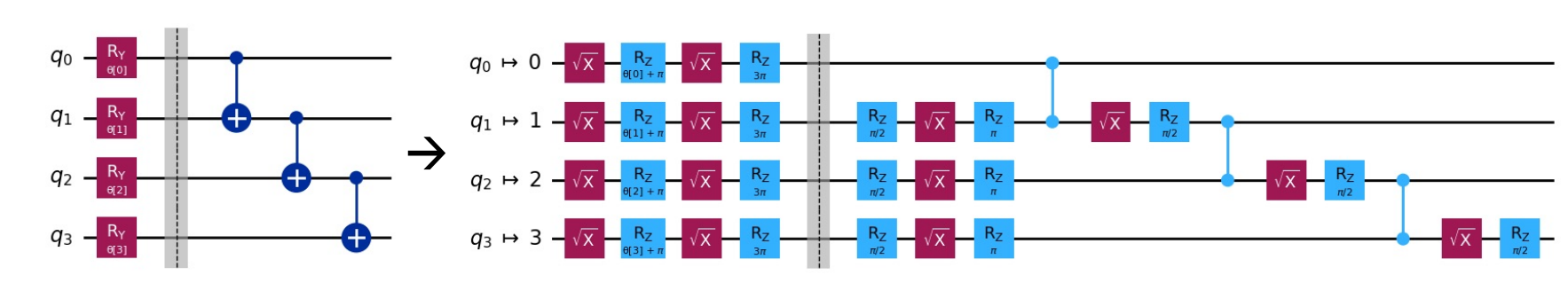}
\caption{Quantum circuit of the real-amplitudes VQE ansatz for $N=4$ qubits with one layer, alongside its transpilation for a linear qubit topology. Each layer has single-qubit $R_y(\theta)$ gates followed by CNOTs between adjacent qubits.}
\label{fig:Circuit}
\end{figure*}

We employ a \texttt{RealAmplitudes} ansatz consisting of single-qubit $R_y(\theta)$ rotations and entangling gates (CNOTs) between neighboring qubits. The \texttt{RealAmplitudes} ansatz provides gate efficiency for PDEs involving real-valued states. It restricts the parameterized single-qubit gates to real-valued rotations, specifically using only $R_y$ gates. Figure~\ref{fig:Circuit} shows a schematic of the $N=4$ version of this ansatz, as well as a sample transpilation for an IBM-like linear qubit layout. Increasing the number of layers in this ansatz increases its expressivity but also its depth and parameter count. The optimizer adjusts the parameters $\{\theta\}$ at each time step, aiming to minimize the VQE cost function that encodes our linear system. In the following, we report the infidelities and resource estimates of this approach, and compare them with those from other quantum dynamics algorithms.

Figure~\ref{fig:infidelities}(a) shows the scalar field $C(x)$ reconstructed at times $t=0,0.25,0.75,1.0$ from the VQE amplitudes for a circuit depth of 10 layers. The solution agrees closely with the DNS ground truth. Figure~\ref{fig:infidelities}(b) plots the infidelity $1-f(t)$ where $f(t) = \lvert\langle C_{\mathrm{DNS}}(t) \mid C_{\mathrm{VQE}}(t)\rangle\rvert^2$. With a converged optimizer the final infidelity drops steeply with depth ($\sim\!10^{-7}$ at $L=3$ and $\sim\!10^{-10}$ at $L=4$), reaching a numerical floor ($\sim\!10^{-14}$) for $L\gtrsim5$ and staying there [Fig.~\ref{fig:infidelities}(c)]. This is expected: The $N=4$ real-amplitudes ansatz at $L=3$ already has $16$ parameters, more than the $15$ needed for an arbitrary real $16$-dimensional state, so the state-vector infidelity is limited only by ansatz expressivity and optimizer or precision tolerance, not by the encoding. All arithmetic is double precision ($\varepsilon\approx2.2\times10^{-16}$); the $\sim\!10^{-14}$ floor therefore lies about two orders of magnitude above $\varepsilon$, set by the optimizer stopping tolerance and round-off accumulated over the $500$ time steps. This
\emph{state-vector} value is therefore an \emph{algorithmic ceiling}: It quantifies how accurately the PDE
can be solved when expectation values are exact, conditional on the chosen ansatz and optimizer rather
than being an absolute bound of the VQLS approach.
Part of why this ceiling is so low is \emph{structural}: By the Proposition of Sec.~\ref{sec:vqe_form} the
ground state of $H_k$ is, by construction, the exact normalized forward-Euler step $(I+\Delta t\hat{A})|C_k\rangle$,
so the linear-solve formulation carries no Trotterization or variational time-integration error, unlike the
dynamics methods (Trotter, VarQTE, and AVQDS) it is benchmarked against, whose noise-free accuracy is additionally
limited by operator splitting or tangent-space time integration. This is a property of the \emph{formulation},
not a claim of accuracy superiority: The infidelity is measured against the \emph{same} discretized DNS that the
solver targets (certifying faithful reproduction of the classical scheme rather than absolute accuracy against
the continuous PDE), and under finite sampling or hardware noise both approaches are limited by the measurement
regime (Sec.~\ref{sec:shotnoise}). The truncation error of the finite-difference DNS relative to the
continuous PDE is the standard $\mathcal{O}(\Delta t,\Delta x^2)$ discretization error and is not
assessed here.
It is \emph{not} a like-for-like comparison with the hardware runs of VarQTE and AVQDS in
Ref.~\cite{alipanah2025quantum} (final infidelities $>10^{-1}$ for $N=4$, $t=1.0$), which are dominated by
hardware noise and finite sampling. To compare like with like, we next quantify the sampling-limited
regime explicitly (Sec.~\ref{sec:shotnoise}), which bridges the state-vector ceiling and the
noisy-hardware results.

Table~\ref{tab:tab1} compares single- and two-qubit gate counts for Trotterization, VarQTE, and AVQDS
(as reported in Ref.~\cite{alipanah2025quantum}) with our VQE ansatz. Two caveats are essential for a fair
reading. First, the counts must be compared at matched ansatz depth: The VarQTE column corresponds to
$L=10$ layers, so we list our VQE ansatz at both $L=5$ and $L=10$. For the real-amplitudes ansatz with linear
entanglement the number of two-qubit gates is exactly $(N-1)L$ (here $3L$ for $N=4$), giving $15$ at
$L=5$ and $30$ at $L=10$; because the ansatz's linear entanglement matches a linear hardware coupling
map, this count incurs no SWAP overhead under transpilation. At the matched $L=10$ it equals VarQTE
($30$) and is fewer than AVQDS ($40$); that is, \emph{comparable}, though not dramatically smaller. Second, and more importantly, the VarQTE and AVQDS circuits
encode the \emph{entire} time integration in a single circuit, whereas the VQE circuit acts on a
\emph{single} time step and must be re-executed $N_{\mathrm{steps}}\approx 500$ times; the VQE
\emph{total} two-qubit count over the full run ($1.5\times10^{4}$ at $L=10$; Table~\ref{tab:totalgates})
therefore exceeds the single final circuit of the variational dynamics methods, though not their full execution cost (see the caption of Table~\ref{tab:totalgates}), and is comparable to a single Trotter circuit. What the table does establish is that the per-step VQE circuit is shallow
(depth $55$ at $L=5$) with a two-qubit count comparable to the variational dynamics methods, while
Trotterization remains far more expensive ($\sim\!10^{5}$ gates). The methods are thus complementary
rather than strictly ordered. For larger PDEs or 2D grids, circuit optimizations (e.g.,
entanglement-preserving layout) or quantum data compression
\cite{gourianov2022quantum, gourianov2022exploiting,gourianov2025tensor} may become essential.

\begin{table}[!htbp]
\caption{\label{tab:tab1}Per-circuit gate counts and depths for $N=4$ qubits in the IBM native basis
$\{X,\sqrt{X},RZ,\textsc{cz}\}$. Trotter, VarQTE, and AVQDS are from \cite{alipanah2025quantum}; the
VarQTE column is at $L=10$ layers and AVQDS at its adaptively converged depth. Our VQE column is the
real-amplitudes ansatz at $L=5$ layers (the depth used in Fig.~\ref{fig:infidelities}). For this ansatz
with linear entanglement the two-qubit count is exactly $(N-1)L=3L$ at $N=4$ ($15$ at $L=5$), with no
SWAP overhead on a linear coupling map; the last row gives the matched $L=10$ two-qubit count ($30$),
equal to VarQTE and fewer than AVQDS. All VQE entries are
\emph{per time step}: The full run repeats the circuit $N_{\mathrm{steps}}\approx500$ times, so the VQE
\emph{total} is not smaller than the dynamics methods (see text).}
\begin{center}
\begin{tabular}{lcccc}
\hline\hline
Gate & Trotter & VarQTE & AVQDS & VQE \\
     &         & ($L{=}10$) &     & ($L{=}5$) \\
\hline
$X$           & 317     & 0       & 6      & 0  \\
$\sqrt{X}$    & 53646   & 108     & 79     & 78 \\
$RZ$          & 48460   & 109     & 67     & 93 \\
$\textsc{cz}$ & 20213   & 30      & 40     & 15 \\
\hline
Total         & 122636  & 247     & 192    & 186 \\
Depth         & 76021   & 90      & 129    & 55 \\
\hline
$\textsc{cz}$ ($L{=}10$) & n/a  & 30  & 40  & 30 \\
\hline\hline
\end{tabular}
\end{center}
\end{table}

\begin{table}[!htbp]
\caption{\label{tab:totalgates}Two-qubit (\textsc{cz}) gate count per circuit and over the full
evolution to $t=1$ ($N=4$). The variational dynamics methods encode the entire evolution in a single
circuit, whereas the VQE circuit is per time step and is repeated $N_{\mathrm{steps}}=500$ times. Per
circuit the VQE is comparable to VarQTE and AVQDS. Both columns count state-preparation gates only: the VQE rows omit the optimizer and measurement circuits of each step, and the VarQTE and AVQDS rows omit the auxiliary Hadamard-test circuits used to estimate the metric and gradient at every step, which Ref.~\cite{alipanah2025quantum} puts at $2046$ per step and $1.023\times10^{6}$ over $t=1$ for $N=4$, $L=10$ (their Appendix~C). Table~\ref{tab:tab1}, the per-circuit count, is the like-for-like comparison.}
\begin{center}
\begin{tabular}{lccc}
\hline\hline
Method & per circuit & circuits & total \\
\hline
Trotter            & $2.0\times10^{4}$ & $1$   & $2.0\times10^{4}$ \\
VarQTE ($L{=}10$)  & $30$              & $1$   & $30$ \\
AVQDS              & $40$              & $1$   & $40$ \\
VQE ($L{=}5$)      & $15$              & $500$ & $7.5\times10^{3}$ \\
VQE ($L{=}10$)     & $30$              & $500$ & $1.5\times10^{4}$ \\
\hline\hline
\end{tabular}
\end{center}
\end{table}

\subsection{Shot-noise baseline: State vector versus finite sampling}
\label{sec:shotnoise}
The results above use exact (state vector) expectation values, which are not directly comparable to
sampling-based results: Alipanah \emph{et al.}'s noiseless emulator already
includes measurement (shot) statistics, and any hardware run necessarily does. We therefore repeat the
VQE time stepping on the \emph{same} noiseless simulator but estimate each circuit's output from a finite
number $S$ of computational-basis measurements, reconstructing the amplitudes as $\sqrt{n_i/S}$ from the
counts $n_i$ (the sign pattern of the real-amplitudes solution is retained). This isolates the
\emph{sampling} contribution to the error, with no hardware noise. It is thus a best-case readout-sampling diagnostic (the solution's signs are assumed known), not a full shot-based VQE optimization, which would also estimate the cost and its gradients from measurements. Recovering the sign structure and, more generally, extracting the amplitude-encoded solution efficiently is the \emph{readout problem} that any quantum differential-equation solver must ultimately confront~\cite{williams2024readout}.

\begin{figure*}[!htbp]
\centering
\includegraphics[width=0.85\linewidth]{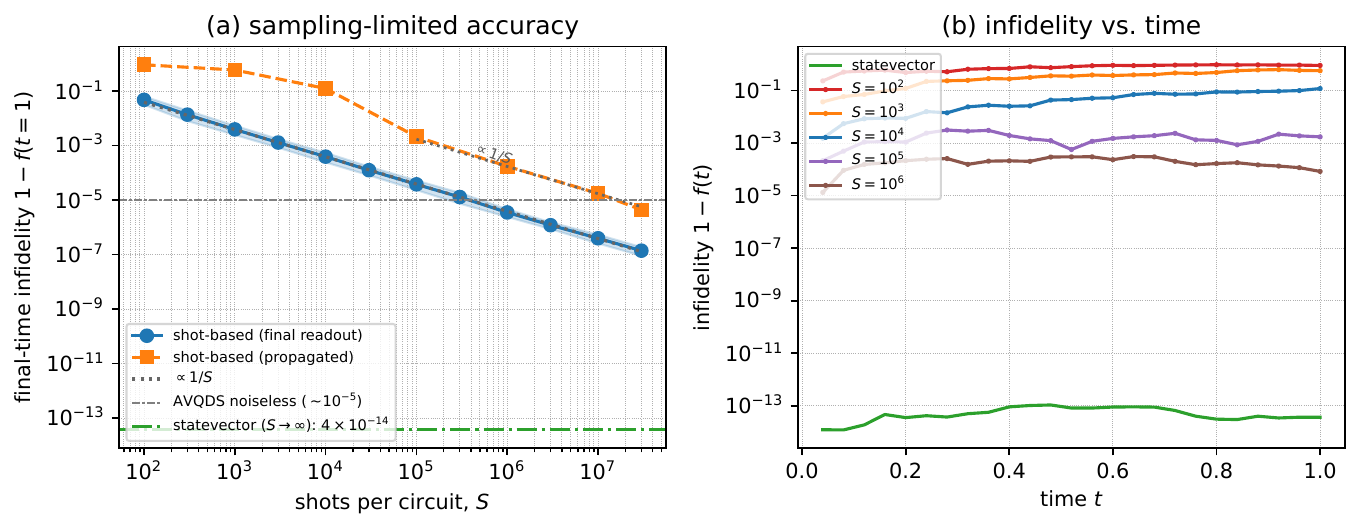}
\caption{\label{fig:shotnoise}Noiseless state vector vs finite-sampling readout diagnostic ($N=4$, $L=10$,
$\Delta t=0.002$). \textbf{(a)} Final-time infidelity vs shots per circuit $S$ (both curves swept to $S=3\times10^{7}$).
Sampling the exact final state (``readout,'' circles) follows the shot-noise scaling $1-f\approx c/S$ (dotted
guide); re-sampling at every time step (``propagated,'' squares) accumulates error over the $500$ steps. The
AVQDS noiseless-emulator level of Ref.~\cite{alipanah2025quantum} ($\sim\!10^{-5}$, dash-dot) and the
state-vector ($S\to\infty$) ceiling are shown for reference: The best-case readout diagnostic reaches this
level near $S\approx4\times10^{5}$ and the realistic propagated sampling near $S\approx2\times10^{7}$; both
are $1/S$-limited with no floor (both dotted guides have slope $-1$), the propagated curve simply carrying
an $\approx\!45\times$ larger shot prefactor from accumulating error over the $500$ steps. \textbf{(b)} Infidelity vs time for the state-vector solution and for propagated shot-based
runs at $S=10^{2}$--$10^{6}$.}
\end{figure*}

Figure~\ref{fig:shotnoise} shows the result for $N=4$, $L=10$. When only the final state is sampled
(``readout''), the infidelity follows a clean shot-noise scaling $1-f\approx c/S$ with $c=\mathcal{O}(1)$ (here $c\approx3$--$5$ over $S=10^{2}$--$3\times10^{7}$, set by the effective number of significant amplitudes in the target state and growing only slowly as diffusion spreads the field or the grid is refined; the slight upward departure from this scaling at the smallest $S$, where $1-f$ is no longer $\ll1$, is the leading correction to the linearization together with the mild positive bias of the $\sqrt{n_i/S}$ amplitude estimate),
falling from $\approx5\times10^{-2}$ at $S=10^{2}$ to $\approx4\times10^{-4}$ at $S=10^{4}$,
$\approx3\times10^{-6}$ at $S=10^{6}$, and $\approx4\times10^{-7}$ at $S=10^{7}$, far above the state-vector
ceiling ($\approx10^{-14}$, the numerical floor of Fig.~\ref{fig:infidelities}); it crosses the AVQDS
noiseless-emulator level ($\sim\!10^{-5}$, Ref.~\cite{alipanah2025quantum}) near $S\approx4\times10^{5}$. When the sampled state
is fed back at \emph{every} step (``propagated,'' the realistic shot-based time-stepping scenario), the
per-step sampling error accumulates over the $500$ steps, so a substantially larger budget is needed:
$S=10^{4}$ yields a final infidelity $\sim\!0.1$, and only $S\gtrsim10^{5}$ ($\sim\!2\times10^{-3}$)
keeps it below $10^{-2}$. The propagated curve nonetheless follows the same $1-f\propto1/S$ law with no
floor once $S$ is large: It reaches $\sim\!2\times10^{-4}$ at $S=10^{6}$ and crosses the AVQDS
noiseless-emulator level ($\sim\!10^{-5}$) near $S\approx2\times10^{7}$ ($\approx4\times10^{-6}$ at
$S=3\times10^{7}$), tracking the readout curve on a parallel $1/S$ line offset by an $\approx\!45\times$
larger shot prefactor (both $1/S$ guides are shown dotted in Fig.~\ref{fig:shotnoise}a), the cost of
accumulating sampling error over the $500$ steps.

This provides a regime-matched comparison with Alipanah \emph{et al.}~\cite{alipanah2025quantum}: Their
noiseless-emulator infidelity ($\sim10^{-5}$) corresponds to a readout shot budget of $S\approx4\times10^{5}$ in
our scaling, whereas their hardware infidelity ($>10^{-1}$) lies above even the $S=10^{2}$ sampling floor
and is therefore dominated by hardware noise rather than by sampling. The state-vector, sampling, and
hardware results thus form a clear hierarchy (algorithmic ceiling $\ll$ sampling floor $\ll$ hardware
error), and it is this decomposition of the error budget, more than any single ``best'' number, that
the benchmark provides.

Our readout scaling $1-f\approx c/S$ may at first appear to differ from the $\propto1/\sqrt{N_s}$ shot-noise
scaling reported for a closely related variational PDE solver, the hybrid quantum-classical Navier-Stokes scheme of Song \emph{et al.}~\cite{song2025incompressible}, whose variational linear-solver cost and
convergence residual bottom out near $1/\sqrt{N_s}$ for $N_s$ shots. The two scalings are in fact consistent,
because they measure different functionals of the same shot statistics. The statistical error of any
\emph{linear} estimator (an amplitude, an expectation value, or a residual norm) scales as $1/\sqrt{S}$
by the central limit theorem. The infidelity, by contrast, is a \emph{quadratic} functional of the state
error: Writing the sampled state as $|\hat\psi\rangle=|\psi\rangle+|\delta\rangle$ with
$\|\delta\|\propto1/\sqrt{S}$, one has
$1-f=1-|\langle\psi|\hat\psi\rangle|^2=\|\delta_\perp\|^2+\mathcal{O}(\|\delta\|^3)\propto(1/\sqrt{S})^2=1/S$,
so $1-f\propto1/S$ is precisely the square of the $1/\sqrt{S}$ amplitude error. We have verified that both
scalings hold simultaneously in our sampled data (the $\ell_2$ amplitude error scales as $1/\sqrt{S}$ and
the infidelity as $1/S$), and the same distinction is visible within Ref.~\cite{song2025incompressible}, whose
$1/\sqrt{N_s}$ residual floor sits well above its reported state infidelity. One caveat: In our study the
shots enter only at \emph{readout} of an exactly optimized state, so $1-f\approx c/S$ is a best-case
bound; a fully shot-based optimization (in which the cost itself is estimated from shots, as in
Ref.~\cite{song2025incompressible}) would additionally incur a $\sim1/\sqrt{S}$ floor on the cost during training.

\subsection{Scaling across problem sizes ($N=4,5,6$)}
\label{sec:Nsweep}
Reference~\cite{alipanah2025quantum} reported results for $N=4,5,6$ qubits (e.g.\ their VarQTE
infidelity-versus-layers study), whereas the experiments above fix $N=4$. The single-register VQLS
formulation of Sec.~\ref{sec:vqe_form} applies unchanged for any $N$: The grid has $2^N$ points and the
VQE acts on $N$ qubits. We repeated the state-vector VQE for $N=5$ ($32$ points) and $N=6$ ($64$ points)
at fixed $\Delta t=0.002$ (the CFL conditions remain satisfied, with
$\mathrm{CFL}_{\mathrm{diff}}=0.016,\,0.064,\,0.256$ for $N=4,5,6$). Table~\ref{tab:Nsweep} shows that the
method scales across the same range as the quantum-dynamics methods: At a fixed, small optimization
budget the infidelity stays at or below the $\sim\!10^{-5}$ noiseless level of
Ref.~\cite{alipanah2025quantum}, and deeper circuits or tighter optimization drive it lower still
($\lesssim10^{-10}$ for $N=4,5$). As expected, larger $N$ requires more layers, since the ansatz must
represent a higher-dimensional state ($2^N-1$ real amplitudes versus $N(L+1)$ parameters), mirroring
the layer-versus-$N$ trend reported for VarQTE in Ref.~\cite{alipanah2025quantum}. This $N$-sweep is a short-horizon scaling check at $t=0.2$, not a full $t=1$ reproduction of Fig.~\ref{fig:infidelities} for every $N$.

\begin{table}[!htbp]
\caption{\label{tab:Nsweep}Statevector VQE infidelity across problem sizes ($Pe=32$, $\Delta t=0.002$,
final time $t=0.2$, small optimizer budget), spanning the $N=4,5,6$ range of
Ref.~\cite{alipanah2025quantum}. The two-qubit count per step is $(N-1)L$.}
\begin{center}
\begin{tabular}{ccccc}
\hline\hline
$N$ & grid $2^N$ & layers $L$ & 2-qubit & $1-f$ \\
\hline
4 & 16 & 10 & 30 & $3\times10^{-11}$ \\
5 & 32 & 10 & 40 & $6\times10^{-11}$ \\
6 & 64 & 15 & 75 & $2\times10^{-5}$ \\
\hline\hline
\end{tabular}
\end{center}
\end{table}

\emph{Reach beyond $N=6$.} The table stops at $N=6$ only to match the range of
Ref.~\cite{alipanah2025quantum}; the construction itself is $N$-agnostic, and the eight-qubit ($2^{8}=256$ amplitude) two-dimensional problem of Sec.~\ref{sec:2d} already exercises it at a larger
register. What grows with $N$ is the \emph{cost}, along three axes. First, our noiseless-state-vector
\emph{verification} manipulates dense $2^{N}\times2^{N}$ operators [$M=(I+\Delta t\hat{A})^{-1}$ and
$H_k$], so it is the classical emulation, not the quantum algorithm, that bounds the present study to
$N\lesssim13$ on a workstation. Second, for the explicit forward-Euler reference at fixed $Pe$ the
diffusive stability number $\mathrm{CFL}_{\mathrm{diff}}=\Delta t\,4^{N}/Pe$ grows with $N$ (here
$0.016,0.064,0.256$ for $N=4,5,6$, and ${>}\tfrac12$ already at $N=7$), so the time step must shrink as
$\Delta t\propto4^{-N}$ (and the step count grow as $4^{N}$) unless an implicit integrator or a larger
$Pe$ is used; this is a property of the classical discretization, independent of the solver. Third, and most
fundamentally, an arbitrary real $2^{N}$-dimensional state carries $2^{N}-1$ degrees of freedom, so a
real-amplitudes ansatz requires $N(L+1)\gtrsim2^{N}-1$, i.e.\ a depth $L\gtrsim2^{N}/N$ that grows
\emph{exponentially} in $N$. The numerical ceiling reached at low depth for $N=4,5$ is thus
special to small registers: The jump to $L=15$ already needed at $N=6$ is the leading edge of this
expressivity wall, beyond which the fixed-depth variational fidelity degrades. This exponential depth
growth is generic to variational state preparation and is precisely why we present the method as a
near-term \emph{benchmark} and error-budget decomposition without claiming a scalable advantage: at
larger $N$ the binding constraint becomes circuit depth, while the formulation is unchanged. Climbing this wall by adding
layers also enlarges the parameter space in which barren plateaus arise, so expressivity and
trainability trade off; ans\"atze that build in the solution's correlation structure, such as
tensor-network (matrix-product or tree) circuits~\cite{gourianov2025tensor,siegl2026tensor}, are a
natural route to parameter efficiency beyond the regime studied here.

\subsection{Two-dimensional demonstration (eight qubits)}
\label{sec:2d}
\begin{figure*}[!htbp]
\centering
\includegraphics[width=\linewidth]{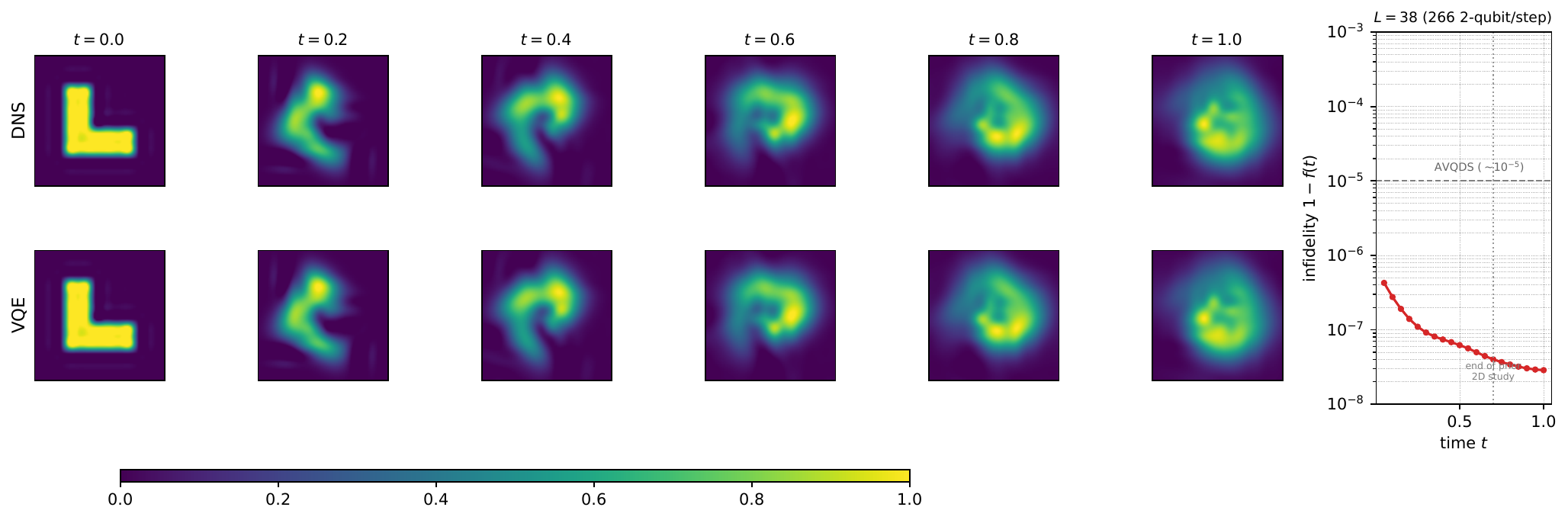}
\caption{\label{fig:2d}Two-dimensional advection-diffusion on eight qubits ($16\times16$ grid, $Pe=132$,
$L$-shaped initial condition, rotational velocity field), matching the 2D case of
Ref.~\cite{alipanah2025quantum}. Top row: Scalar field $C(x,y)$ from DNS at $t=0$ to $t=1$ (covering and
exceeding the $t\le0.7$ of Ref.~\cite{alipanah2025quantum}'s 2D contour figure); bottom row:
the single-register VQLS-VQE solution, in essentially exact visual agreement as the profile rotates
and diffuses (fields shown with periodic cubic interpolation). Right: Infidelity $1-f(t)$ for the
depth-$38$ ansatz ($266$ two-qubit gates per step), at the exact-expectation algorithmic-ceiling level
(worst $\sim\!7\times10^{-7}$ within the first few steps), decreasing monotonically to
$2.9\times10^{-8}$ by $t=1$; the dashed
line marks the sampling-limited $\sim\!10^{-5}$ of the corresponding
\emph{shot-based} 2D AVQDS simulation in Ref.~\cite{alipanah2025quantum} (a different measurement regime,
cf.\ Sec.~\ref{sec:shotnoise}).}
\end{figure*}

Reference~\cite{alipanah2025quantum} also treated a two-dimensional advection-diffusion problem with
eight qubits ($16\times16$ grid). We apply this single-register VQLS-VQE to the 2D case using their
setup: a rotational velocity field $\mathbf{U}=(-y,x)/\sqrt{x^2+y^2}$, P\'eclet number $Pe=132$,
periodic boundaries, and an $L$-shaped initial profile. The $256$-cell field on the cell-centered grid
over $[-\tfrac12,\tfrac12)^2$ (a half-cell offset avoids the $r=0$ rotation singularity) is flattened as
$k=i\,m+j$ into a $2^{8}$-amplitude vector. The generator $\hat A=\hat A_{\mathrm{diff}}+\hat A_{\mathrm{adv}}$
combines a diffusion term ($\Gamma=1/Pe$ times the standard five-point Laplacian, i.e.\ a Kronecker sum
$D_2\!\oplus\!D_2$ of the one-dimensional second-difference operator) with a central-difference
discretization of the rotational advection $\mathbf{U}=(-y,x)/\sqrt{x^2+y^2}$; because the advection
coefficients vary with position, $\hat A_{\mathrm{adv}}$ is not a constant-coefficient Kronecker product
but is assembled from the same one-dimensional stencils applied along each axis. No change to the
formulation is required; the same VQE acts on $2N=8$ qubits. Figure~\ref{fig:2d} shows that the VQE solution reproduces the
rotating-and-diffusing profile in essentially exact visual agreement with the DNS. With a sufficiently
expressive ansatz (depth $38$, with $266$ two-qubit gates per step) and \emph{exact} (state-vector)
optimization, the infidelity stays at or below the $\sim\!10^{-7}$ level over the \emph{entire} evolution
to $t=1$, the same final time simulated by Ref.~\cite{alipanah2025quantum} (whose 2D contour figure,
their Fig.~7, is shown only to $t=0.7$). The infidelity in fact \emph{decreases monotonically} with time as
diffusion smooths the field, from a worst case of $7\times10^{-7}$ early in the evolution (within the
first few steps, where the per-step field change is largest) to $2.9\times10^{-8}$ at $t=1$.
This $\sim\!10^{-7}$ is the exact-expectation \emph{algorithmic ceiling}; the $\sim\!10^{-5}$ reported for
the 2D AVQDS simulation of Ref.~\cite{alipanah2025quantum} is a shot-based (noiseless-emulator) value, so,
as in the 1D analysis of Sec.~\ref{sec:shotnoise}, the roughly two-order gap is a \emph{measurement-regime} effect, with no algorithmic advantage implied. The point is that the ground-state solver extends
directly to multidimensional problems and attains, in the noise-free regime, the same low
algorithmic-ceiling infidelity in 2D as in 1D.

\emph{Resource comparison.} For near-term (NISQ) hardware the binding cost is the \emph{per-circuit}
two-qubit gate count, since each circuit's error grows with its own gate count and the circuit must fit
within the device coherence time; the number of distinct circuits executed is a throughput concern, not a
single-circuit feasibility one. Our VQE uses a \emph{fixed} \texttt{RealAmplitudes} ansatz ($L=38$,
$N(L+1)=312$ parameters and $(N-1)L=266$ two-qubit gates). Alipanah \emph{et al.}'s 2D AVQDS instead grows an
\emph{adaptive} ansatz from a weight-$3$ pool of $848$ real-preserving Pauli strings~\cite{alipanah2025quantum}. Running their
released code for this case (Table~\ref{tab:2dres}), the ansatz reaches $\gtrsim700$ two-qubit gates early
in the evolution and keeps growing, so \emph{per circuit} our fixed ansatz is the leaner of the two by
nearly a factor of three ($266$ versus $788$ in our run), consistent with AVQDS being the leanest of the \emph{dynamics} methods
(Trotter, VarQTE, and AVQDS) of Ref.~\cite{alipanah2025quantum}, yet still a several-hundred-gate circuit at eight qubits. The 1D case shows the same ordering (Table~\ref{tab:tab1}: VQE $15$--$30$ versus AVQDS $40$
two-qubit gates). The AVQDS count is adaptive and grows with the simulated horizon, so the quoted value is
a lower bound, still growing; our $266$-gate circuit reaches the exact-expectation ceiling whereas AVQDS
tracks the dynamics at infidelity $\sim\!10^{-5}$.

\begin{table}[!htbp]
\caption{\label{tab:2dres}Per-circuit two-qubit-gate count for the 2D (eight-qubit, $16\times16$) problem,
the NISQ-relevant resource, since each circuit's error scales with its own gate count. The AVQDS entry
is from running the released code of Ref.~\cite{alipanah2025quantum} (weight-$3$ pool of $848$ operators);
its adaptive ansatz \emph{grows} with the simulated horizon, so its parameter count and its $\gtrsim700$
two-qubit gates are lower bounds, reaching $788$ in our run and still growing. ``VQE'' is this work (fixed ansatz,
$N(L{+}1)=312$ parameters).}
\begin{center}
\begin{tabular}{lcc}
\hline\hline
 & VQE & AVQDS~\cite{alipanah2025quantum} \\
\hline
qubits          & $8$             & $8$ \\
ansatz          & fixed, $L{=}38$ & adaptive ($w{=}3$) \\
parameters      & $312$           & grows \\
2-qubit gates   & $266$           & $\gtrsim700$ ($788$) \\
\hline\hline
\end{tabular}
\end{center}
\end{table}

\emph{Implementation caveat.} In this work $H_k$ is constructed classically and its expectation values are
evaluated either exactly (state vector) or through the readout-sampling diagnostic of
Sec.~\ref{sec:shotnoise}. A direct hardware realization would additionally require an efficient
decomposition or block-encoding of $\mathcal{A}$ (and hence $H_k$), preparation of $|b\rangle=|C_k\rangle$,
and measurement of the resulting terms. We quantify the Pauli decomposition of $\mathcal{A}$ for the
$N=4$ operator in Appendix~\ref{app:pauli} ($23$ Pauli strings, or equivalently a three-term combination
of the identity and two cyclic-shift unitaries), leaving its conditioning to future work, and treat the
present study as a controlled benchmark of ansatz expressivity and error budgets, short of a
complete hardware protocol.

\section{Conclusion and Outlook}
\label{sec:Conclusions}

We have benchmarked a VQE approach to solving the 1D advection-diffusion PDE, comparing it to the Trotter and variational real and imaginary time-evolution methods of Ref.~\cite{alipanah2025quantum}. Although those methods were implemented on IBM hardware and subject to significant noise, we used an idealized state-vector simulator to assess the \emph{intrinsic} performance of a VQE-based linear-systems solver. Numerical results for $N=4$ qubits demonstrate final-time infidelities at a numerical floor ($\sim\!10^{-14}$) once the depth reaches $L\approx5$, with only tens of two-qubit gates per time step. We also delimit the scope: We do not derive real-time Hamiltonian-simulation error bounds, and we do not optimize the ansatz structure beyond the fixed RealAmplitudes form; both are natural extensions, beyond the scope of this work.

In this paper, we focus on a \emph{state-vector} simulation of a VQE-based 
approach, providing a noise-free baseline for algorithmic performance. 
While our results are comparable to Alipanah \emph{et al.}'s
noiseless emulator outcomes, they differ in two key respects~\cite{alipanah2025quantum}. First, our state-vector VQE bypasses any sampling error by computing exact expectation values from the full wave function, whereas Alipanah \emph{et al.}'s shot-based emulator includes measurement statistics (albeit with no hardware noise).
Section~\ref{sec:shotnoise} quantifies this directly: With $S$ shots per circuit the infidelity is sampling-limited to $\approx c/S$, so the noiseless-emulator value of Alipanah \emph{et al.}\ ($\sim10^{-5}$) corresponds to $S\sim10^{5}$, and the state-vector result is the $S\to\infty$ ceiling. Second, we solve the PDE via a VQE linear-system formulation instead of simulating real or imaginary-time evolution. Both approaches are valid but differ in circuit structure and required gates. Hence, direct comparisons in overall gate counts or final errors must account for these structural differences.

Nevertheless, comparing our noise-free (state-vector) VQE results to 
the noise-free shot-based emulator, and ultimately to their noisy 
hardware runs, helps disentangle purely algorithmic performance from 
the hardware's limitations. Our final infidelities, down to the
numerical floor ($\sim\!10^{-14}$) in the \emph{idealized} scenario, illustrate the potential headroom
before noise becomes dominant. Moving forward, exploring shot noise 
and explicit noise models (e.g.\ amplitude damping or depolarizing 
channels) will bridge the gap between idealized results and the 
realities of NISQ hardware performance.

In a noise-free setting, the VQE converges to the discretized PDE solution at moderate circuit depth, demonstrating the algorithmic capability of variational PDE solvers when hardware error is not the limiting factor, consistent with recent VQE benchmarks for spin models on near-term hardware \cite{li2023benchmarking}. Trotterization remains gate-intensive ($\sim 10^5$) for $N=4$, whereas VQE is comparable in gate count to VarQTE or AVQDS while reaching numerical-floor accuracy in this idealized setting. For linear PDEs, VQE-based approaches are thus competitive in per-circuit depth for the small instances studied here. Alipanah \emph{et al.} did not isolate or statistically characterize shot noise, and their final solutions were degraded by hardware errors~\cite{alipanah2025quantum}; our state-vector results establish the algorithmic ideal against which current hardware performance can be measured.

In future work, we plan to incorporate hardware-like noise models (e.g., amplitude damping, depolarizing channels, readout error) into VQE PDE simulations and measure the effect on fidelity.
Building on the two-dimensional eight-qubit demonstration of Sec.~\ref{sec:2d}, we will further scale to larger 2D and 3D PDEs relevant for incompressible flows and finance (e.g., Refs.~\cite{lipton2001mathematical, herman2023quantum, lipton2024hydrodynamics}). The persistent challenge is to maintain accuracy while controlling circuit depth and mitigating NISQ noise, possibly through error mitigation \cite{PhysRevLett.121.220502,PhysRevA.94.052325}, dynamical decoupling \cite{PhysRevA.58.2733,Das2021adapt,ozguler2024qudit}, or specialized ans\"atze \cite{li2017efficient, grimsley2019adaptive}. 

Overall, our results indicate that near-term quantum hardware will continue to face noise-related limitations, but from an \emph{algorithmic} standpoint, VQE is a viable approach for small PDE grids, complementary to existing quantum-dynamics methods, with a comparable per-step circuit (though a larger total over the full time integration; see Sec.~\ref{sec:sec3-1}). As hardware matures, combining VQE-like strategies with advanced error mitigation, data compression, and adaptive circuit design could open a path toward a clearer assessment of when quantum PDE solvers may become practically advantageous.

\section*{Acknowledgments}
The author acknowledges discussions with the authors of Ref.~\cite{alipanah2025quantum} and the support from the University of Pittsburgh and Pittsburgh Quantum Institute. The author acknowledges valuable discussions at the 2024 Quantum Industry Day, held at the Simons Institute for the Theory of Computing in Berkeley, California, and the AI + Quantum Workshop at the Aspen Center for Physics in February 2025. This work was supported in part by the Aspen Center for Physics, which is supported by the National Science Foundation Grant No.~PHY-2210452.

\section*{Data availability}
The code and data that support the findings of this article, including scripts that regenerate every figure and table, are openly available in Zenodo~\cite{ozguler2026punq}.

\appendix

\section{Equivalent block formulation and a worked $N=4$ example}
\label{app:formulation}

\emph{Block (register-doubling) form.}\quad
The single-register system \eqref{eq:linsys} can equivalently be written by carrying both the current
and next states in one vector. Stacking the data constraint and the Euler relation gives the
$2^{N+1}\times2^{N+1}$ block-lower-triangular system
\begin{equation}
  \mathcal{A}_{\mathrm{blk}}\,|x\rangle = |b\rangle,
  \quad
  \mathcal{A}_{\mathrm{blk}}=
  \begin{pmatrix} I & 0 \\ -(I+\Delta t\hat{A}) & I \end{pmatrix},
  \label{eq:blk}
\end{equation}
with $|x\rangle=(|C_k\rangle;\,|C_{k+1}\rangle)$ and $|b\rangle=(|C_k\rangle;\,0)$. The top block
enforces the first half of $|x\rangle$ to equal $|C_k\rangle$, and the bottom block enforces
$|C_{k+1}\rangle=(I+\Delta t\hat{A})|C_k\rangle$. Solving
\eqref{eq:blk} as the ground state of $\mathcal{A}_{\mathrm{blk}}^\dagger(I-|b\rangle\langle b|)
\mathcal{A}_{\mathrm{blk}}$ reproduces the same evolution as \eqref{eq:Hk}, but on $N+1$ qubits. Because
the top block of $|x\rangle$ merely repeats the known input $|C_k\rangle$, the doubling is unnecessary;
the single-register form \eqref{eq:linsys}--\eqref{eq:Hk} removes it, and all results in this paper use
$N$ qubits.

\emph{Worked $N=4$ example.}\quad
With $N=4$ ($m=16$, $\Delta x=1/16$, $Pe=32$), the coefficients of \eqref{eq:eq4} are
$b=-2/\Delta x=-32$, $c=1/\Delta x-Pe/2=16-16=0$, and $d=1/\Delta x+Pe/2=32$, with prefactor
$1/(Pe\,\Delta x)=1/2$. Hence $\hat A$ is the $16\times16$ circulant with diagonal $-16$, sub-diagonal
$+16$, and vanishing super-diagonal (with periodic wrap $\hat A_{0,15}=16$): At this P\'eclet number the
central scheme reduces to upwind advection. With $\Delta t=0.002$,
$(I+\Delta t\hat A)$ has diagonal $0.968$ and sub-diagonal $0.032$. Starting from the box state
\eqref{eq:ic}, $|b\rangle=|C_0\rangle=\tfrac1{\sqrt8}(0,0,0,0,1,1,1,1,1,1,1,1,0,0,0,0)^\top$, the first
step forms $H_0=M^\dagger(I-|C_0\rangle\langle C_0|)M$ with $M=(I+\Delta t\hat A)^{-1}$, and the VQE
ground state of $H_0$ is $|C_1\rangle\propto(I+\Delta t\hat A)|C_0\rangle$. Its nonzero amplitudes (on
grid points $i=4,\dots,12$), after normalization, are $0.344$ at $i=4$, $0.355$ at $i=5,\dots,11$, and
$0.011$ at $i=12$ (zero elsewhere): The box advected by one upwind step, with the trailing edge at
$i=4$ dropping and a small leading amplitude appearing at $i=12$. The circuit operates on $N=4$ qubits ($16$ amplitudes), the same problem
size as the $N=4$ case of Ref.~\cite{alipanah2025quantum}; iterating $H_k\!\to\!|C_{k+1}\rangle$ for
$k=0,\dots,499$ gives the full evolution.

\section{Pauli decomposition of the advection-diffusion operator}
\label{app:pauli}
A hardware realization must express the operator $\hat A$ (equivalently $I+\Delta t\hat A$, which shares
its off-diagonal structure) in a measurable basis. Expanding the $N=4$, $Pe=32$ operator in the Pauli
basis reproduces it exactly with $23$ nonzero strings of the $256$ possible. Their weights (the
number of non-identity factors) are distributed as one weight-$0$ term (the identity), two weight-$1$,
four weight-$2$, eight weight-$3$, and eight weight-$4$, so the maximum weight equals the register size;
the full-weight terms originate in the periodic wraparound of the discretization. The count grows as
$\tfrac32\,2^{N}-1$ ($23$, $47$, and $95$ for $N=4,5,6$) with maximum weight $N$, so term-by-term Pauli
measurement is not scalable on its own.

The operator is, however, far more compact as a linear combination of unitaries. From the finite-difference
stencil,
\begin{equation}
  \hat A \;=\; d_0\,I \;+\; c_{+}\,U_{+} \;+\; c_{-}\,U_{-},
  \label{eq:Alcu}
\end{equation}
with constant coefficients $d_0=-2/(Pe\,\Delta x^2)$ and $c_{\pm}=1/(Pe\,\Delta x^2)\mp 1/(2\Delta x)$, and
$U_{\pm}$ the two cyclic shift-by-one (increment and decrement) permutation operators, which are unitary and
admit $O(N)$-gate circuit implementations (for example via QFT-based adders). A hardware variational
linear-solver would build the cost from this three-term structure instead of the exponential Pauli
expansion; the inverse $M=(I+\Delta t\hat A)^{-1}$ is dense and is never formed explicitly.

\bibliographystyle{apsrev4-2}
\bibliography{refs} 

\end{document}